\documentstyle[aps,prl,epsfig,multicol]{revtex}
\begin{document}
\title{Capture and release of a conditional state of
a cavity QED system by quantum feedback.}
\author{W. P. Smith${}^{1}$, J. E. Reiner${}^{1}$, L. A.
Orozco${}^{1}$, S. Kuhr${}^{2}$, and H. M. Wiseman${}^{3}$}
\address{${}^{1}$Dept. Physics and Astronomy, SUNY Stony
Brook, Stony Brook NY 11794-3800, USA.\\ ${}^{2}$ Institut f{\"u}r
Angewandte Physik, Universit{\"a}t Bonn, Wegelerstr. 8, D-53115 Bonn, Germany.\\
${}^{3}$ Center for Quantum Dynamics, School of Science, Griffith
University, Brisbane, Queensland 4111, Australia.}
\date{\today}
\maketitle

\begin{abstract}
Detection of a single photon escaping an optical cavity QED system
prepares a non-classical state of the electromagnetic field. The
evolution of the state can be modified by changing the drive of
the cavity. For the appropriate feedback, the conditional state
can be captured (stabilized) and then released. This is observed
by a conditional intensity measurement that shows suppression of
vacuum Rabi oscillations for the length of the feedback pulse and
their subsequent return.
\end{abstract}
\begin{multicols}{2}
Feedback control of quantum systems was first studied
 about fifteen years ago \cite{YamImoMac86,HauYam86,Sha87},
 in the field of quantum optics.
In these approaches, the feedback could be understood in an
essentially classical way, with quantum field theory entering only
to dictate the magnitude of the fluctuations. This is possible if
fluctuations are small compared to the mean fields being detected.
More recently, a different approach to quantum optical feedback
has been developed \cite{WisMil93b,Wis94a}, based on quantum
trajectories \cite{Car93b,DumZolRit92,MolCasDal93}, which specify
the stochastic evolution of a quantum state conditioned on
continuous monitoring (such as by photodetection). This theory
allows the treatment of feedback in the deep quantum regime, where
quantum fluctuations are not small compared to the mean. It is
also arguably the best way to approach feedback, as the
conditioned state by definition comprises all of the knowledge of
the experimenter on which feedback could be based
\cite{DohJac99,Doh00}.

So far, experiments in quantum feedback, such as Refs.~%
\cite{YamImoMac86,WalJak85b,TapRarSat88,MerHeiFab91,Tau95,Buc99},
have all been in the
regime of small fluctuations \cite{mit}.
Cavity QED is able to explore the
opposite regime, where fluctuations in the conditional state are
large. Furthermore, using the theory of quantum trajectories,
Carmichael and coworkers \cite{carmichael00,foster00} showed that
such conditional quantum fluctuations are intrinsically related to
the production of squeezing and antibunching. In this letter we
present experimental results for the application of feedback in
this regime. Following a photodetection, the conditioned quantum
state of the system is $|\psi_{\rm c}(\tau)\rangle$. Given our
knowledge of this evolution, we can, for certain times $\tau$,
change the parameters of the system dynamics so as to capture the
system in that conditioned state. When the parameters are later
restored to their usual values, the released system state resumes
its interrupted evolution. This directly demonstrates both the
reality of the conditioned state and its usefulness for quantum
feedback.

A Cavity quantum electrodynamical (QED) system consists of a
single mode of the electromagnetic field of a cavity interacting
with one or a collection of $N$ two-level atoms \cite{berman94}.
Microwave Cavity QED systems have been used recently to prepare
multiparticle entanglement \cite{rauschenbeutel00}, and to produce
photon number states of the electromagnetic field \cite{varcoe00}.
Operated at optical frequencies, cavity QED systems can now trap
single atoms in the electric field of the cavity when its average
occupation is about one photon \cite{hood00,pinkse00}. The system
size and dynamics are characterized by two dimensionless numbers:
The saturation photon number $n_{0}$ and the single atom
cooperativity $C_{1}$. They scale the influence of a photon and
the influence of an atom in cavity QED. These two numbers relate
the reversible dipole coupling of a single atom with the cavity
mode ($g$) with the irreversible coupling to the reservoirs
through cavity ($\kappa$) and atomic radiative ($\gamma$) decays:
$C_{1}=g^{2}/\kappa \gamma$ and $n_{0}=\gamma^{2}/3g^{2}$. The
strong coupling regime of cavity QED requires $n_{0} \leq 1$, and
$C_{1} \geq 1$. Strictly speaking, the coupling constant ($g$) is
spatially dependent and together with the $N$ moving atoms may be
described by effective constants.

With weak driving, the system can be accurately modelled as having
either zero, one, or two excitations of the coupled normal modes
of the field and the atoms. In this regime, photodetections are
very infrequent and the state before a detection can be taken to
be the steady state, which is almost pure:
\begin{eqnarray}
|\psi_{\rm ss}\rangle &=& |0,G\rangle\ + \lambda\left(|1,G\rangle -
\frac{2g\sqrt{N}}{\gamma}|0,E\rangle\right)\nonumber\\ && +\,
\lambda^2\left(\zeta_0\frac{1}{\sqrt{2}}|2,G\rangle -
\theta_0\frac{2g\sqrt{N}}{\gamma}|1,E\rangle\right) + \cdots.
\label{psiss}
\end{eqnarray}
\noindent Here $|n,G\rangle$ represents $n$ photons with all $(N)$
atoms in their ground state, $|n,E\rangle$ represents $n$ photons
with one atom in the excited state with the rest $(N-1)$ in their
ground state. The small parameter is
 $\lambda=\langle\hat a \rangle = \epsilon /\kappa
(1+C_1N)$, which depends on the input driving field $\epsilon$,
while $\zeta_0$  and $\theta_0 $ are coefficients of order unity
for the two-excitation component of the state that can give rise
to a photon detection, 
and depend on $g$, $\kappa$, and $\gamma$
\cite{carmichael91,reiner01}. After the photodetection occurs
$|\psi_{\rm ss}\rangle$ collapses to $\hat a|\psi_{\rm ss}\rangle
/ \sqrt{{\langle \hat a^{\dagger} \hat a \rangle}_{\rm ss}}$. This
evolves as the conditioned state:
\begin{eqnarray}
|\psi_{\rm c}(\tau)\rangle &=& |0,G\rangle\ +
\lambda\left(\zeta(\tau)|1,G\rangle -
\theta(\tau)\frac{2g\sqrt{N}}{\gamma}|0,E\rangle\right)\nonumber\\ &&
+\, O(\lambda^2) \label{psiconditioned}
\end{eqnarray}
This is different from the initial state because $\zeta$ (the
``field'' evolution) and $\theta$ (the ``atomic polarization''
evolution) oscillate coherently at the vacuum Rabi frequency
$(g\sqrt{N}$) over time as the system re-equilibrates exchanging
energy between the atomic polarization and the cavity field
\cite{carmichael91,reiner01}.

If we choose a time $\tau=T$ for Eq.~(\ref{psiconditioned}) such
that $\zeta(T)=\theta(T)$ then, to order $\lambda$ we obtain
\begin{equation}
|\psi_{\rm c}(T)\rangle \simeq |0,G\rangle\ + \lambda'\left(|1,G\rangle
- \frac{2g\sqrt{N}}{\gamma}|0,E\rangle \right)
,\label{condition1}
\end{equation}
This  is of the form of $|\psi_{\rm ss}\rangle$ in
Eq.~(\ref{psiss}) but with a different mean field
$\lambda'=\zeta(T)\lambda.$ This conditional state can be
stabilized if, at time $T$, we change the driving amplitude by a
factor $\zeta(T)$. Given the almost $90^{\circ}$ out of phase
oscillations between the field $(\zeta)$ and the atomic
polarization $(\theta)$ \cite{reiner01} the time $T$ is  close to
the time when the field is crossing zero.

Conditional quantum states such as Eq.~\ref{psiconditioned} can be
measured using high order quantum optical correlations
\cite{foster00,mandel95}. When the light transmitted through the
cavity (with annihilation operator $\hat{a}$) is split the photons
enter two detectors. The normalized correlation function of the
two photocurrents is the time-and normally ordered average
\begin{eqnarray}
g^{(2)}(\tau) &=& \frac{\langle \hat a^{ \dagger} (t) \hat
a^{\dagger} (t+\tau) \hat a (t + \tau) \hat a (t)
\rangle_{\rm ss}}{\langle \hat a^{\dagger} (t) \hat a (t)
\rangle^{2}_{\rm ss}} \nonumber \\
&=& \frac{\langle\hat n (t+\tau) \rangle_{\rm c} }
{\langle \hat n(t)\rangle_{\rm ss}}, \label{sec}
\end{eqnarray}
where $\hat{n}=\hat{a}^{\dagger}\hat{a}$, and
c means ``conditioned on a detection at time $t$ in steady
state''. If a detection at one detector is used to trigger a feedback
pulse on the system,  the correlation function will no longer be time
symmetric. However, for $\tau > 0$ the expression (\ref{sec})
still measures the conditional state in the presence of feedback:
\begin{equation}
g^{(2)}(\tau) \simeq \frac{|\langle 1,G|\psi_{\rm c}(\tau) \rangle|^2}{|
\langle 1,G|\psi_{\rm ss}|^2}=[\zeta(\tau)]^2\label{eqg2cond}
\end{equation}

\begin{figure}
\leavevmode \centering \epsfxsize=3.15in \epsffile{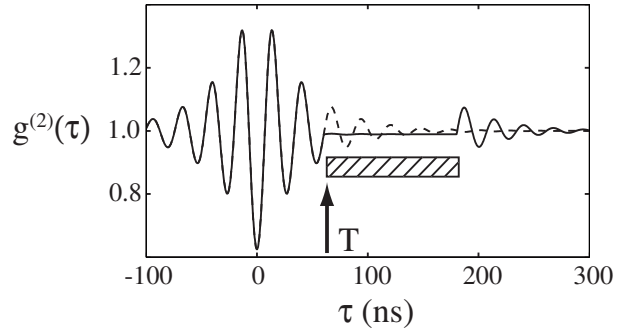}
\caption {\narrowtext Conditional evolution of cavity QED system
with (continuous line) and without (dashed line) feedback. The
shaded rectangle region corresponds to an instantaneous step down
of 0.2 \% of the driving intensity applied at time $T$.
\label{figure1}}
\end{figure}

Fig. \ref{figure1} shows the conditional evolution of the state of
the cavity QED system, as given by Eq. \ref{eqg2cond}. We start
with the quantum theory valid for $N$ two level atoms identically
coupled to the cavity in the weak field regime
\cite{carmichael91}. We find $g_{\rm {eff}} < g$ and $N_{{\rm
eff}}$ \cite{rempe91,carmichael99} using our experimentally
determined values for $g^{(2)}(0)$ such that $g\sqrt{N}=g_{\rm
{eff}}\sqrt{N_{{\rm eff}}}$. All broadening effects are
incorporated by the modification of the atomic decay rate, $\gamma
\rightarrow \gamma'$. We numerically solve the time evolution with
the driving step incorporated. This simplified approach agrees
with our previous work for $g^{(2)}(\tau)$ \cite{fosterpra00}. The
dashed line is the free evolution of the system, and shows the
time symmetry of the correlation function. The application of a
feedback pulse at time $T$ alters the evolution of the system. The
continuous lines shows the evolution when the step change in the
driving intensity $\epsilon$ satisfies the conditions necessary to
reach a new steady state described by Eq. \ref{condition1}. The
parameters of the calculation are those of our experiment:
$g\sqrt{N}/2\pi=37.3$ MHz, $\gamma'/2\pi=9.1$ MHz and
$\kappa/2\pi=3.7$ MHz. The change in the intensity is small (0.2
\%) and here we assume that its rise and fall are instantaneous.
The new state reached by the system after the change of driving
intensity no longer shows the vacuum Rabi oscillations and instead
has a new value for the steady state slightly lower than the
original one. The duration of the pulse that changes the steady
state is finite and our model shows the reappearance of the
oscillation delayed by the length of the pulse.

Our cavity QED apparatus, described in detail in Ref.
\cite{fosterpra00}, consists of the cavity, the atomic beam, an
excitation laser, and the detector system. Two high reflectivity
curved mirrors (input transmission mirror 10 ppm, output
transmission mirror 285 ppm, and separation $l=880\mu$m) form the
optical cavity (waist of the TEM$_{00}$ mode $w_0=34\mu$m). A
Pound-Drever-Hall stabilization technique keeps the cavity locked
to the appropriate atomic resonance. An effusive oven produces a
thermal (440 K) beam of Rb atoms with an angular spread of 2.8
mrads at the cavity mode. A laser beam intersects the atomic beam
before the atoms enter the cavity in a region with 2.5 Gauss
uniform magnetic field. It optically pumps all the $^{85}$Rb atoms
of the $F=3$ ground state the magnetic sublevel $F=3, m_F=3$. The
three rates that characterize our cavity QED system are
$(g,\kappa,\gamma/2)/2\pi = (5.1,3.7,3.0)$ MHz.

\begin{figure}
\leavevmode \centering \epsfxsize=3.15in \epsffile{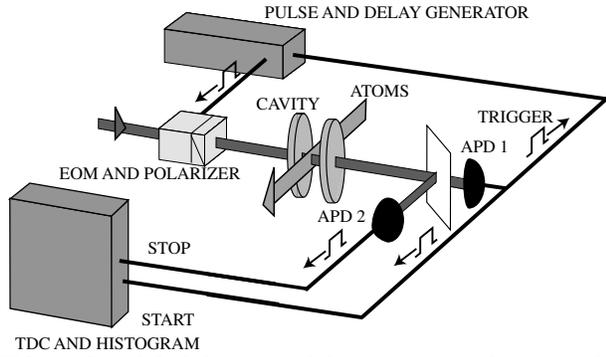}
\caption{\narrowtext Simplified diagram of the experimental setup.
The output of the cavity QED system passes through a beam splitter
(entrance of an intensity correlator) such that the detection of a
photon at the ``start" avalanche photodiode (APD) also triggers a
change in the driving intensity trough a pulse that drives an
electro-optical modulator EOM in front of a polarizer. A histogram
of the  delays between the ``start'' and ``stop'' gives the
conditional evolution of the intensity \label{figure2}}
\end{figure}

Fig. \ref{figure2} shows a schematic of our apparatus. Light from
a Ti:Sapph, locked to the $5S_{1/2}, F=3 \rightarrow 5P_{3/2},
F=4$ transition of $^{85}$Rb at 780 nm, drives the cavity QED
system. The signal escaping the cavity creates photodetections at
the ``Start''and ``Stop'' avalanche photodiodes (APD). The output
pulse of the ``Start'' detector is split and one part sent to the
start channel of the correlator [time to digital converter (TDC)
with 0.5 ns per bin, histograming memory, and computer] while the
other goes to a variable time delay, and after pulse shaping and
lengthening, drives an Electro Optical Modulator (EOM) in front of
a polarizer to produce a pulse of 8 ns risetime and 120 ns length
in the driving intensity of the cavity. The delay between the
detection of a photon at APD1 and the arrival of the pulse at the
cavity can be as short as 45 ns. The other APD sends its pulses to
the correlator to stop the TDC that measures the time interval
between the two events.

We operate the cavity QED system in a non-classical regime where
the size of the vacuum Rabi oscillations is large enough to permit
their rapid identification during data taking. We begin by
measuring the antibunched second order correlation function of the
intensity escaping our cavity QED system. We then apply the step
change in the driving intensity at time $T$ to fulfill the
conditions of Eq. (\ref{condition1}) and obtain a new steady
state.

\begin{figure}
\leavevmode \centering \epsfxsize=3.15in \epsffile{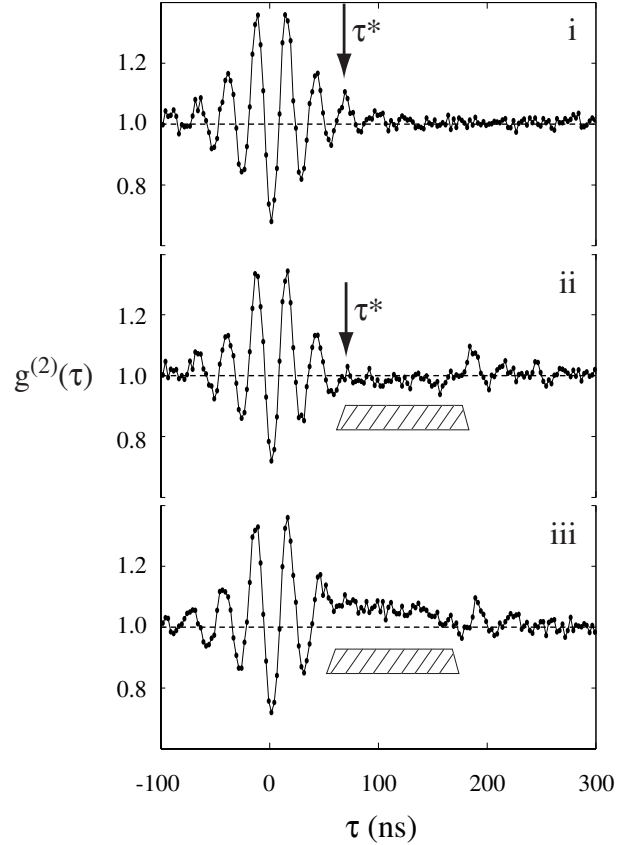}
\caption{\narrowtext Measured intensity correlation function with
the feedback step applied during the shaded region: i) no feedback
($g\sqrt{N}/2\pi=37$ MHz), ii) suppression with a step down change
of 2.6 \% ($g\sqrt{N}/2\pi=37$ MHz)  iii) suppression with a step
up change of 3.9\% ($g\sqrt{N}/2\pi=31$ MHz). The oscillation of
the system continues with the same phase and amplitude once the
step is off. Note that the time $T$ for the beginning of the
feedback in ii and iii is different as indicated by the position
of the shaded region. The data has been binned in 2.5 ns points
joined with a line.\label{figure3}}
\end{figure}

Fig. \ref{figure3} shows measurements of the correlation function
in the absence (i) and presence (ii, iii) of feedback. Traces i
and ii have the same oscillating frequency while for trace iii we
have a smaller number of atoms. $\tau^*$ marks the position where
the oscillation we want to suppress reaches a maximum.  The steady
state for $\tau$ large corresponds to an intracavity intensity of
$n/n_0=0.07$. Fig. \ref{figure3} ii shows the correlation function
with step down feedback (- 2.6 \%) for 120 ns, beginning at
$\tau=T=57$ ns, when the oscillation crosses unity and is growing.
The oscillation that has a maximum in trace i at the point marked
by $\tau^*$ has disappeared, the steady state is lower that marked
by the dashed line, and the oscillation reappears after the pulse
is turned off with approximately the same size of the amplitude as
the suppressed one. Trace iii shows step up feedback (+ 3.9 \%) at
$T=46$ ns when the phase is opposite from trace ii.

Reversing the sign of the step produces an enhancement of the
oscillations. If the time $T$ for the application of the pulse is
not correct, it is not possible to achieve good suppression. There
is qualitative agreement between the traces i and ii with those of
the theory  in Fig. \ref{figure1}. They show the suppression and
the delayed return of the oscillation by the application of a
feedback pulse to the driving intensity. Although the theoretical
model does not include all the experimental details that give rise
to broadening the main features of the response are clearly
explained.

\begin{figure}
\leavevmode \centering \epsfxsize=3.15in \epsffile{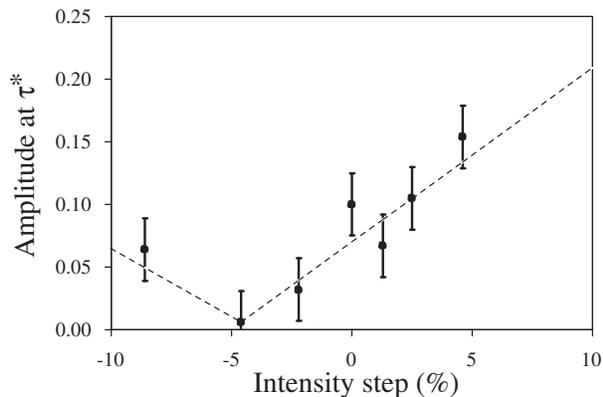}
\caption {\narrowtext Amplitude of the normalized intensity
response at time $\tau*$ (first oscillation extreme of
$g^{(2)}(\tau)$ after the application of the feedback pulse) as a
function of the size the feedback step. The dashed line is a
theoretical prediction. \label{figure4}}
\end{figure}

We have followed the size of the amplitude of the oscillation
immediately after we apply the feedback pulse, at the time
$\tau^*$ defined in Fig. \ref{figure3}, to make a quantitative
comparison with theory. Fig. \ref{figure4} shows the results for a
series of measurements that include steps up (positive) and steps
down (negative). The theory (dashed line) incorporates the
measured shape of the pulse (at the point -4.6\%), all sources of
dephasing present in the system are modelled by the polarization
decay rate $\gamma'/2\pi= 9.1$ MHz. The plot shows both
enhancement and suppression with quantitative agreement.

The quantum feedback in this system is triggered by a fluctuation
(detection of a photon) that is large enough to significantly
modify the system, because of the strong coupling. This detection
prepares the system in an evolving conditional state. We then
change the drive of the system and are able to freeze its
evolution to a new time-independent steady state. The suppressed
oscillations return once the pulse turns off, with the same  phase
and amplitude information. This sort of quantum feedback is a
novel way to manipulate the fragile conditional states that come
from strongly coupled systems.

We would like to thank J. Gripp and J. Wang for their interest and
help with this project. This work has been supported by the NSF,
NIST, DAAD, and the Australian Research Council.

\end{multicols}
\end{document}